\documentclass[prx,twocolumn,superscriptaddress,preprintnumbers]{revtex4} 

\usepackage{graphicx}
\usepackage{bm}
\usepackage{amsmath}
\usepackage{amssymb}
\usepackage{amsfonts}
\usepackage{latexsym,epsfig,bbm}

\newcommand{\bra}[1]{\langle{#1}|}
\newcommand{\ket}[1]{|{#1}\rangle}

\newcommand{\figref}[1]{Fig.~\ref{#1}}

\newcommand{\Tr}{\mathrm{Tr}}
\newcommand{\PP}{\mathcal{P}}
\newcommand{\PQ}{\mathcal{Q}}
\newcommand{\ident}{\mathbb{I}}

\usepackage{color}
\definecolor{orange}{rgb}{.8,.4,.1}
\definecolor{purple}{rgb}{.7,.1,.7}

\begin{document}

\title{Quantum Quasi-Zeno Dynamics: Transitions mediated by frequent projective measurements near the Zeno regime}

\author{T. J. Elliott}
\email{thomas.elliott@physics.ox.ac.uk}
\affiliation{Department of Physics, Clarendon Laboratory, University of Oxford, Parks Road, Oxford OX1 3PU, United Kingdom}
\author{V. Vedral}
\affiliation{Department of Physics, Clarendon Laboratory, University of Oxford, Parks Road, Oxford OX1 3PU, United Kingdom}
\affiliation{Centre for Quantum Technologies, National University of Singapore, Singapore 117543}
\affiliation{Department of Physics, National University of Singapore, 2 Science Drive 3, 117551 Singapore}
\affiliation{Center for Quantum Information, Institute for Interdisciplinary Information Sciences, Tsinghua University, 100084 Beijing, China}

\date{\today}

\begin{abstract}
Frequent observation of a quantum system leads to quantum Zeno physics, where the system evolution is constrained to states commensurate with the measurement outcome. We show that, more generally, the system can evolve between such states through higher-order virtual processes that pass through states outside the measurement subspace. We derive effective Hamiltonians to describe this evolution, and the dependence on the time between measurements. We demonstrate application of this phenomena to prototypical quantum many-body system examples, spin chains and atoms in optical lattices, where it facilitates correlated dynamical effects.
\end{abstract}
\maketitle

\section{Introduction}
Reminiscent of the arrow paradox put forth by Zeno of Elea, concerning the apparent discrepancy in the motion of objects when they can at any and all instants be observed to be stationary, the quantum Zeno effect (QZE) \cite{misra1977, neumann1955mathematical, teuscher2004} argues that the act of continuously observing a quantum state leads to a zero probability of evolving away from the state, thus freezing the system evolution. This effect has been extended to encompass the case of degenerate measurement subspaces, where multiple states of the system possess identical outcomes of the measured observable, such that evolution within this subspace is unhindered by the measurement, a phenomenon called quantum Zeno dynamics (QZD) \cite{facchi2000, facchi2002, facchi2008}. The QZE and QZD have been observed in a range of experimental setups, including ions \cite{itano1990}, photons \cite{kwiat1995}, nuclear magnetic resonance spins \cite{xiao2006}, atoms in microwave cavities \cite{signoles2014}, Bose-Einstein condensates \cite{streed2006,schafer2014}, and Rydberg atoms \cite{patil2015}. There has also been much theoretical interest in the field, particularly because of the opportunities offered by measurement-based control of a system \cite{nakazato2003, nakazato2004, erez2004, wu2004, compagno2004, militello2004, wang2008, erez2008, pazsilva2012, raimond2012, burgarth2013, everest2014, stannigel2014, elliott2015a, mazzucchi2015, elliott2015b}.

It has been shown that even when consecutive measurements are finitely spaced, the locking to a measurement subspace can still occur \cite{layden2015}. However, in this case, the description of the system evolution solely in terms of this subspace is incomplete \cite{dhar2015}; the finite time between measurements allows higher-order processes to occur, where the system first transitions away from the measurement subspace, and then subsequently back in to it before the next measurement, thus preserving the value of the measured observable. Similar effects have been predicted for continuous measurement in the quantum jump formalism \cite{mazzucchi2015, kozlowski2015}.

In this article we demonstrate how these higher-order processes, which we call quantum quasi-Zeno dynamics (QqZD), arise from perturbative considerations of standard QZD. We find effective Hamiltonians to describe the evolution of the system, and suggest interpreting such processes as virtual transitions, providing a simple illustrative example using a three-level system. We extend the formalism to encompass time-dependent Hamiltonians, non-equally- and stochastically-spaced measurements, and discuss how transitions to different measurement subspaces may be incorporated into the treatment. We then apply this formalism to exhibit how this phenomenon may manifest in two archetypal examples of many-body systems, spin chains and atoms in optical lattices, where we show that the higher-order processes correspond to correlated dynamics in the system.

\section{Fundamentals of Quantum quasi-Zeno Dynamics}
Consider a system evolving under Hamiltonian $H$. Between measurements, the evolution of the quantum state $\rho$ after time $t$ is described by the unitary operator $U(t)=\exp(-iHt)$,  through $\rho\to U\rho U^\dagger$ \cite{dirac1967principles} (we use natural units $\hbar=1$). This system is subject to measurement from an external source, and we model the effect of a measurement of the observable $A=\sum_j A_j \PP_j$ with outcome $A_k$ to modify the state according to $\rho\to \PP_k \rho \PP_k$, where $\PP_k$ is the projector for the subspace containing all states with measurement eigenvalue $A_k$ (see Appendix \ref{secderivation}).

In this formalism, we can describe the evolution of a system subject to frequent measurement. For two consecutive measurements a time $\delta t$ apart, a state $\rho$ initially in eigenspace $\PP$ of the measurement operator evolves $\rho\to\PP'U(\delta t)\rho U^\dagger(\delta t)\PP'$, where $\PP'$ is the subspace of the measurement outcome. In the limit where $H\delta t$ is small, we can expand the exponential $U(\delta t)\approx 1-iH\delta t -H^2\delta t^2/2+\mathcal{O}(\delta t^3)$, to calculate the probability that the measurement outcome changes. The probability that the measurement results in a value corresponding to subspace $\PQ\neq\PP$ is hence given by $P(\PQ)=\Tr(H \rho H \PQ)\delta t^2+\mathcal{O}(\delta t^3).$  Summing this over all measurement subspaces different to $\PP$, we have the condition that for the probability of a change in the measurement value to occur to be negligible, we require $\sum_{\PQ} P(\PQ)\ll 1$, i.e.~for any state $\rho$ in $\PP$, $\sum_{\PQ} \mathrm{Tr} (\PQ H \rho H) \delta t^2 \ll 1$. This forms our `Zeno-locking' requirement on timescales for the periods between measurements. From here, we assume this condition is met.

After $N\gg1$ such measurements in a time $\tau=N\delta t$, each resulting in the same measurement value, with subspace $\PP$, the system evolution can be approximated by the effective evolution operator $U_{\mathrm{eff}}(\tau)=\exp(-i H_{\mathrm{eff}} \tau)$ (see Appendix \ref{secderivation} for details) , with the corresponding effective Hamiltonian 
\begin{equation}
\label{eqHeff}
H_{\mathrm{eff}}=\sum_{k=1}^\infty \frac{(-i\delta t)^{k-1}}{k!}H_Z^{(k)},
\end{equation}
where we define the quasi-Zeno Hamiltonians $H_Z^{(k)}=\PP H ((\ident - \PP)H)^{k-1} \PP$. In the limit that $\delta t\to 0$, i.e.~the standard QZD scenario, this evolution becomes $\exp(-i H_Z^{(1)} \tau)$, with $H_Z^{(1)}$ being the standard Zeno Hamiltonian \cite{facchi2008}, recovering the QZD result. However, when $\delta t$ is small-but-finite, we have the more general QqZD scenario, where the quasi-Zeno Hamiltonians $H_Z^{(k)}$ mediate $k$th-order transitions where the initial and final states are in the measurement subspace $\PP$, but intermediate states are not. Because of the dependence of each quasi-Zeno Hamiltonian's contribution to the effective Hamiltonian on increasing powers of $\delta t$, each one is less significant than that of the previous order, and the intimate dependence of QqZD on the measurement timestep is evident. Indeed, as the probability of a measurement outcome belonging to a different subspace scales as $\delta t^2$, in practice it is likely that the second-order quasi-Zeno Hamiltonian $H_Z^{(2)}$, also scaling as $\delta t^2$ forms the only significant deviation from standard QZD, with the higher-order $H_Z^{(k)}$ forming corrections on top of this. 

Heuristically, we can see that since both the transition probability and the second-order quasi-Zeno Hamiltonian have similar magnitude $\sim\mathrm{Tr}(\PP H\PQ H \PP\rho)\delta t^2$, the QqZD correction within a given subspace is of the same order as the probability to transition out of the subspace Thus, provided we are in the regime for which QqZD is valid (i.e.~this quantity is much less than unity), the relative magnitude of the correction to the state at the time at which a transition ultimately occurs is approximately independent of $\delta t$. The size of $\delta t$ is still relevant however, as it governs the accuracy of the approximate effective evolution operator (more accurate for smaller $\delta t$), the size of the correction due to the higher-order quasi-Zeno Hamiltonians (decreasing with $\delta t$), and the total timescales over which the QqZD evolution takes place (longer for smaller $\delta t$). Note that the standard Zeno dynamics takes place on timescales independent of $\delta t$. Because the QqZD correction is of a similar magnitude to the transition probability, the quasi-Zeno deviation can become very non-negligible especially when one considers long experimental runs, such as those where the measurement subspace changes during the trajectory (see Section \ref{secfurther}).

In the standard QZD regime, the effective Hamiltonian is simply the Zeno Hamiltonian $H_Z^{(1)}$, which, being Hermitian, leads to unitary dynamics. Contrastingly, the quasi-Zeno Hamiltonians are alternatively Hermitian and anti-Hermitian, and thus due to the second-order quasi-Zeno Hamiltonian $H_Z^{(2)}$ being non-vanishing for any non-trivial Hamiltonian and measurement operator when $\delta t$ is finite, in the quasi-Zeno regime the dynamics is non-unitary. Instead, the dynamics of the system will tend towards the eigenstate(s) of $H_Z^{(2)}$ with lowest eigenvalue that can be accessed by the dynamics from the (quasi-)Zeno Hamiltonians, and the decay in the norm of the state corresponds to the survival probability of remaining in the measurement subspace.

From the above, we have that the probability of a transition out of the measurement subspace between times $t$ and $t+\delta t$ is given by
\begin{align*}
P(\bar{\PP},t)&=\mathrm{Tr}(H\rho(t)H(\ident-\PP))\delta t^2+\mathcal{O}(\delta t^3)\\&\approx\mathrm{Tr}(H_Z^{(2)}\rho(t))\delta t^2.
\end{align*}
Thus, the total survival probability of remaining in the initial measurement subspace after $N$ measurements is given by $\prod_{n=1}^N (1-P(\bar{\PP},n\delta t)$. For long times, the survival probability will tend to zero, unless there is a state space which satisfies Tr($H_Z^{(2)}\rho)=0$. The second-order quasi-Zeno evolution causes the system to tend towards this state, effecting a `natural selection' of states, removing those for which the survival probability is lower, tending towards a steady state space. More generally, the evolution tends towards effective steady states which minimise the rate of higher-order processes the system undergoes, and hence those with the largest survival probability. Such effective steady states are fragile, as they have a non-zero transition probability, and so for longer times will eventually transition out of the measurement subspace.

\section{Interpretation and Example}
\label{secinterp}
Physically, the QqZD second-order terms involve a small-but-finite occupation of an intermediate state between measurements, which is then removed by the projection of the subsequent measurement, provided the locking of the measurement eigenvalue is maintained. While this intermediate state is occupied, it can transition to other states as usual for the non-measurement case. These transitions can either be to states also outside of the measurement subspace (in which case occupation of these states is also removed by the next measurement), or back in to the measurement subspace, but not necessarily into the original state. Higher-order terms involve transitions with additional intermediate states. When the time between measurements decreases, the maximum occupation of the intermediate states will also be decreased, and hence the rate of transitions out of these states will be lessened. This is reflected in the form of the effective Hamiltonians and their dependence on $\delta t$. In the infinitely frequent measurement limit of QZD, there is no occupation of the intermediate states, and thus there are no transitions beyond first-order. 

Because of the locking to the measurement subspace, occupation of the intermediate states is never directly observed. However, through the occurrence of transitions that take place via these states, their temporary occupation may be indirectly inferred. As a result of this, and because the dynamics can be described by effective Hamiltonians acting only on the measurement subspace, allowing a description of the intermediate states to be omitted, these states outside of the measurement subspace can be viewed as virtual states, and consequently, that the transitions between states in the measurement subspace that pass through these virtual states can be seen as virtual processes. Similar transitions via such virtual states are also present in the continuous, non-projective measurement case \cite{mazzucchi2015, kozlowski2015}, where they are compared with Raman-like processes.

\begin{figure}
\centering
\includegraphics[width=\linewidth]{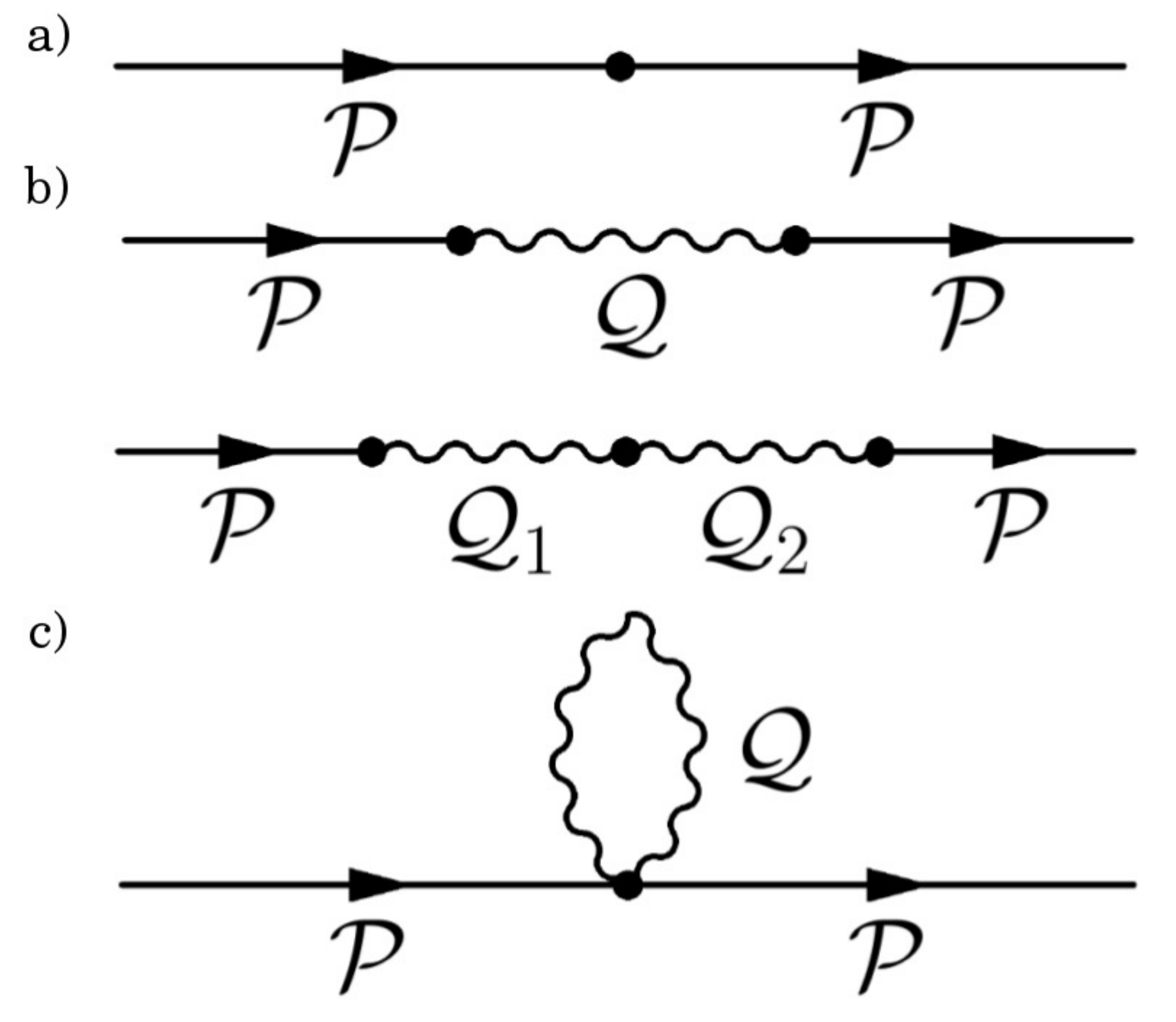}
\caption{{\bf Transitions via virtual processes}: The quasi-Zeno Hamiltonians give rise to transitions that occur via states outside the measurement subspace, which can be viewed as virtual processes. These can be compared conceptually with Feynman diagrams, where instead of interactions occurring via virtual particles, we instead represent transitions occurring via occupation of virtual states. Single vertex processes (a) are mediated by the standard Zeno Hamiltonian $H_Z^{(1)}$, while higher-order processes (b) with $n$ vertices and $n-1$ virtual states are mediated by the quasi-Zeno Hamiltonian $H_Z^{(n)}$. Transitions to a virtual state and back to the initial state can be represented by loop diagrams (c), akin to the representation of self-interaction with Feynman diagrams.}
\label{figfeynman}
\end{figure}

We draw visual analogy with Feynman diagrams \cite{zee2010} for these virtual processes (see \figref{figfeynman}). Feynman diagrams, used as pictorial representations of interactions in high-energy physics, depict interactions as being mediated by virtual particles. We can construct a similar picture for the virtual processes of QqZD, where the incoming and outgoing lines are the initial and final states in the measurement subspace, the vertices are the transitions between states, and the internal lines correspond to occupation of the intermediate states. In this representation, the number of vertices corresponds to the number of transitions, and hence the order of the quasi-Zeno process; transitions described by the Zeno Hamiltonian $H_Z^{(1)}$ have one vertex and no virtual states, as they do not require occupation of the intermediate states, whilst transitions from the second-order quasi-Zeno Hamiltonian $H_Z^{(2)}$ are represented by two vertices and one virtual state. Second-order processes that return back to the same initial state can be considered akin to self-interacting processes, giving rise to self-energy type contributions to the Hamiltonian. We note however, that this analogy is intended as a graphical aid to interpret the transitions within the QqZD framework, and we are not proposing that the mathematics of the processes described by Feynman diagrams be directly mapped onto QqZD.

To further illustrate these virtual processes, we use a very basic toy model consisting of the simplest system that can exhibit non-trivial QqZD: a three-state system where two states possess a degenerate measurement eigenvalue. Consider such a system, say a spin-1 particle, with states $\{\ket{-1},\ket{0},\ket{1}\}$, where the label signifies the $S^Z$ value of the state. The particle is subject to a transverse field of strength $\lambda$, such that it has Hamiltonian $H=\lambda S^X=(\lambda/\sqrt{2})(\ket{-1}\bra{0}+\ket{0}\bra{1}+h.c.)$, and frequent measurement is made of the magnitude of its spin value ($A=|S^Z|$). The corresponding projectors for the measurement subspaces are $\PP_0=\ket{0}\bra{0}$ for $A=0$, and $\PP_1=\mathbb{I}-\PP_0=\ket{-1}\bra{-1}+\ket{1}\bra{1}$ for $A=1$.

The Hamiltonian contains no direct transitions between the $\ket{-1}$ and $\ket{1}$ states, instead requiring the state to first go via the $\ket{0}$ state. Thus, in such a setup in the standard QZD scenario, the Zeno Hamiltonian vanishes for both measurement subspaces; $H_Z^{(1)}=0$. However, when the measurements are finitely frequently spaced, as in the QqZD regime presented here, the second-order quasi-Zeno Hamiltonian for the $A=1$ subspace is non-zero; applying the appropriate projectors to obtain the second-order quasi-Zeno Hamiltonian, we find for the $\{\ket{-1},\ket{1}\}$ subspace that it takes the form
\begin{equation}
\label{eqhsimple}
H_Z^{(2)}=\frac{\lambda^2}{2}(\ket{-1}\bra{-1}+\ket{1}\bra{1}+\ket{1}\bra{-1}+\ket{-1}\bra{1}),
\end{equation}
where the first two terms are `self-energy' type contributions from the system transitioning to $\ket{0}$ and back in to the initial state, while the latter terms give rise to transitions between the $|S^Z|=1$ states, again by sequentially undergoing two transitions to and from state $\ket{0}$. This intermediate state $\ket{0}$ is never observed to be occupied, and thus the transitions appear as a virtual processes. 

This simple example can be straightforwardly solved to find the steady state to which QqZD drives the system. The eigenvalues of Eq.~\eqref{eqhsimple} are 0 and $\lambda^2$, with associated eigenstates $\ket{-}=(\ket{-1}-\ket{1})/\sqrt{2}$ and $\ket{+}=(\ket{-1}+\ket{1})/\sqrt{2}$ respectively. Thus, according to standard QZD, a system initialised in the state $\ket{-1}=(\ket{+}+\ket{-})/\sqrt{2}$ subject to such a measurement will remain in this state, while in contrast, QqZD predicts the system evolution to be $(\mathrm{exp}(-\lambda^2t\delta t/2)\ket{+}+\ket{-})/\sqrt{2}$, the decay in the norm representing the probability to remain in the measurement subspace. Thus, according to QqZD, the long-term evolution of the system is towards the state $\ket{-}$, provided the system remains in the same measurement subspace, which occurs with probability 1/2. Interestingly, this final state is a dark state of the original Hamiltonian $(H\ket{-}=0)$, and hence once this steady state is reached, the system will remain in it even if the measurement is no longer performed.

As three-level systems are routinely realised in a variety of experimental setups, this example may also provide a useful schematic for an initial experimental demonstration of QqZD.

\section{Further Generalisations}
\label{secfurther}
In the above, we took the time between measurements to be equal for simplicity. Generalisation to non-equal timesteps between measurements is straightforward. The form of the quasi-Zeno Hamiltonians are unchanged, but the dependence on $\delta t$ now leads to differing strengths of the $H_Z$ between each measurement in the effective Hamiltonian. We can modify the effective evolution to account for this by including a product over effective evolutions for all the different measurement timesteps. Taking $\delta t_j$ as the time between measurements $j-1$ and $j$, with $\sum_{j=1}^N\delta t_j=\tau$, this can be written
$$U_{\mathrm{eff}}(\tau)=\prod_{j=1}^Ne^{\sum_{k=1}^\infty \frac{(-i\delta t_j)^{k}}{k!}H_Z^{(k)}}.$$
Considering only the terms up to $\mathcal{O}(\delta t)$ in the total evolution, we can neglect those arising from the non-commutativity of effective Hamiltonians for differing timesteps, as they occur at higher-order, suppressed by a factor $\delta t_j-\delta t_{j'}$, and hence approximate $U_{\mathrm{eff}}(\tau)=\exp(-iH_Z^{(1)}\tau-\sum_{j=1}^N H_Z^{(2)}\delta t_j^2/2)$.

With the evolution written explicitly in terms of each measurement timestep, we can also clearly see how time-dependent Hamiltonians may be incorporated into the formalism, at least for cases where they can be treated as being approximately piecewise constant between measurements, by generalising the quasi-Zeno Hamiltonians $H_Z^{(k)}(t)=\PP H(t)((\mathbb{I}-\mathcal{P}H(t))^{k-1}\PP$ in the above effective evolution, and imposing appropriate time-ordering. This generalisation also allows for systems where the measurement timestep depends on a stochastic process (for example, the decay of a particle) to be described. If the variance in the timesteps for such a process is sufficiently narrow, an `average' trajectory could be considered by calculating the moments $\langle\delta t^n\rangle$, with an average number of measurements $\langle N\rangle = \tau/\langle \delta t\rangle$.

Thus far, we have taken the measurements to occur sufficiently close together that the measurement outcome can be assumed to be constant. However, it is possible to relax this condition, still with measurement occurring much more frequently than changes to the measured value, and describe the system evolution by a straightforward extension to the QqZD formalism. Between changes in the measurement value, the system is described by the appropriate QqZD effective evolution operator for this subspace. When the measurement eigenvalue changes, from a value corresponding to subspace with projector $\PP$ to that of subspace with projector $\PP'$, the change in the state is $\rho\to\PP'H\PP\rho\PP H\PP'$, from the leading term $\mathcal{O}(\delta t^2)$ allowing transitions out of the measurement subspace. After the measurement, the system is again described by a QqZD effective evolution operator, but now that corresponding to the new subspace. In an experimental run, one can simply determine when this change in subspace occurs by observing when the measurement value changes. 

\section{Application to Many-Body Systems}
Many-body systems often possess interesting properties that are described by observables dependent on the collective state of multiple particles. Different configurations of particles can still result in the same system-wide measurement value for this observable; such configurations hence correspond to the same measurement subspace. This makes many-body systems a suitable arena for QqZD, and we shall here provide examples of many-body systems, along with associated observables formed of linear functions of the occupation numbers of the system modes, showing how this can lead to correlated dynamics.

\subsection{Spin Chains}

\begin{figure}
\centering
\includegraphics[width=\linewidth]{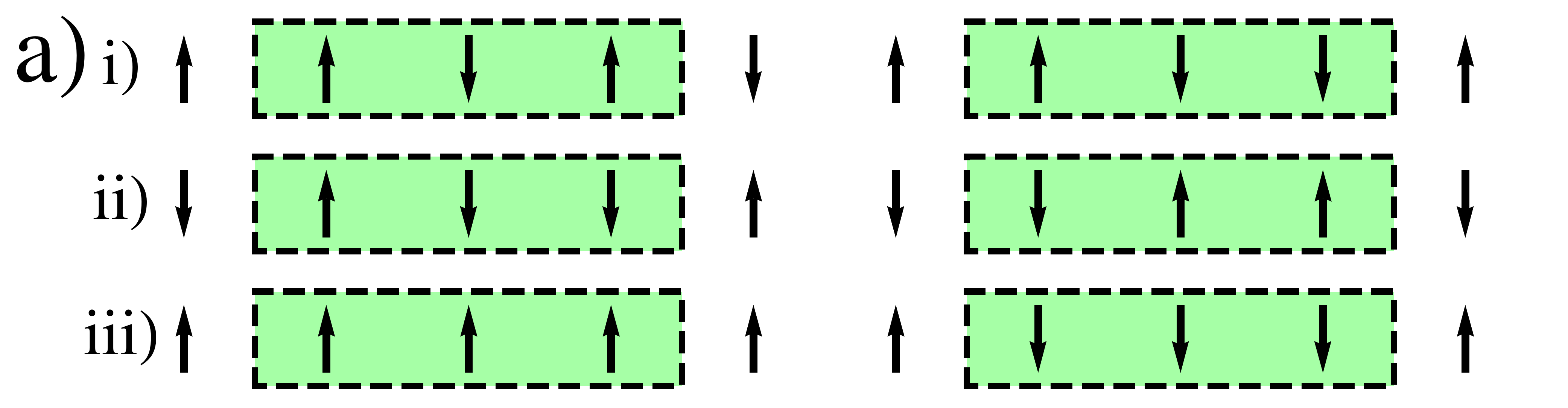}
\includegraphics[width=\linewidth]{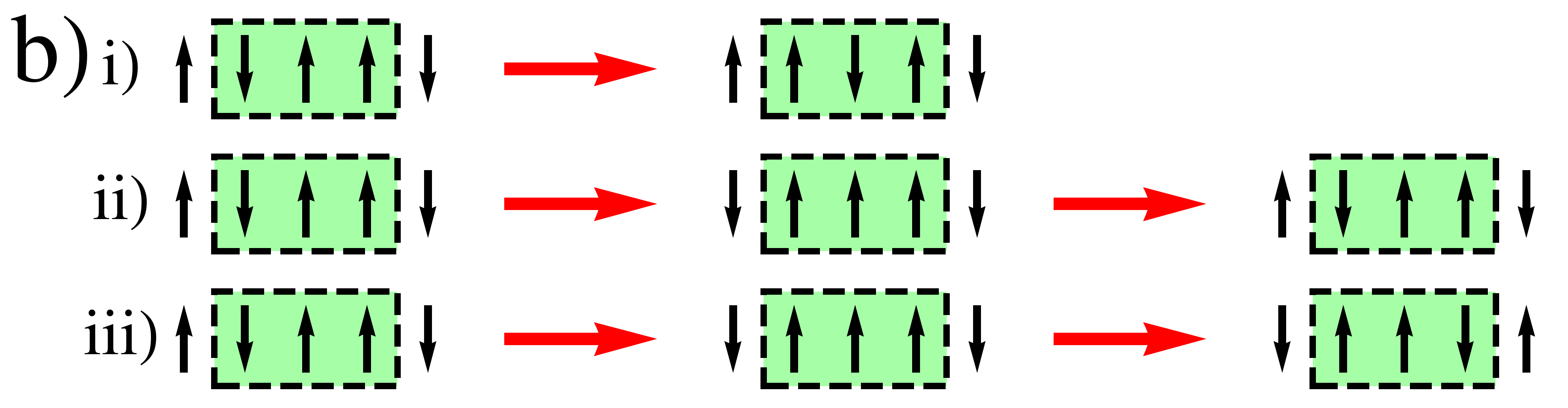}
\caption{{\bf Correlated processes in spin chains}: Multiple configurations of spins (a) belong to the same measurement subspace. Processes that ultimately preserve the measurement can take place, at (b$i$) first- or (b$ii$,$iii$) higher-order. Here, the measurement is signified by the total magnetisation of the green regions.}
\label{figspins}
\end{figure}

For the first example, we consider an array of spins in a chain \cite{mattis2006theory},  where each pair of neighbouring spins is coupled by an exchange interaction $S_i^+ S_{i+1}^-$, such that the full system Hamiltonian is
$$H=-J\sum_{\langle i j \rangle} S^+_iS^-_j$$
where $J$ is the coupling strength of the interaction and $\langle ij\rangle$ indicates that $i$ and $j$ are spins on neighbouring sites. Further terms that can be added to this Hamiltonian which we do not consider here in our examples are biases due to external fields/anisotropies $\lambda S^{X,Y,Z}$ and spin-spin interactions along the $Z$ axis $\tilde{J}S^Z_iS^Z_j$. We take the total magnetisation (the sum of $S^Z$ values) of a set of sites as our measurement, such that the subspaces are defined by states with the same magnetisation in this region [\figref{figspins}(a)]. Dynamics changing this magnetisation are forbidden by the Zeno-locking, and thus the standard Zeno Hamiltonian contains  only spin-exchange between neighbouring spins with either both, or neither, in the measured regions. That is, we can write the standard Zeno Hamiltonian as
\begin{equation}
\label{eqspinsone}
H_Z^{(1)}=-J(\sum_{\langle i\in\mathcal{A},j\in\mathcal{A}\rangle}S^+_iS^-_j+\sum_{\langle i\in\mathcal{B},j\in\mathcal{B}\rangle}S^+_iS^-_j),
\end{equation}
where $\mathcal{A}$ is the set of sites in the measured region(s), and $\mathcal{B}$ the unmeasured sites.

However, the higher-order quasi-Zeno Hamiltonians mediate correlated spin-exchange events, where multiple pairs of spins flip approximately simultaneously between measurements, conserving the total magnetisation measured. The second-order quasi-Zeno Hamiltonian can be written
\begin{equation}
\label{eqspinstwo}
H_Z^{(2)}=J^2\sum_{\substack{\langle i\in\mathcal{A},j\in\mathcal{B}\rangle\\ \langle k\in\mathcal{A},l\in\mathcal{B}\rangle}}(S^+_iS^-_jS^-_kS^+_l+S^-_iS^+_jS^+_kS^-_l).
\end{equation}
These processes mediate two such correlated exchanges, involving only pairs that straddle the measurement region boundaries, and can be of two forms: in the first, both exchanges happen between the same pair (i.e.~$i=k$ and $j=l$), but in opposite directions, thus leaving the individual spins unchanged; in the second, the two pairs are distinct, with one exchange increasing the total magnetisation of the measurement region, while the other decreases it. These processes are illustrated in \figref{figspins}(b). In the latter case, there is no restriction on the spatial separation of the two pairs, and hence these processes can be correlated over long distances; this then resembles a superexchange interaction \cite{kramers1934, anderson1950}, but with the potential for longer separation between the pairs.

\subsection{Atoms in Optical Lattices}

Analogous processes can be considered for atoms in an optical lattice. In state-of-the-art setups, lattices containing bosonic atoms have been generated inside optical cavities \cite{landig2015, klinder2015}, and these setups allow for linear functions of the atomic occupation of each site to be measured through the leakage of light from the cavity after having been scattered by the atoms \cite{mekhov2007, elliott2015a}. In the absence of measurement, and with negligible cavity backaction, the atoms behave according to the Bose-Hubbard Hamiltonian \cite{jaksch1998} $H=-J\sum_{\langle ij\rangle}b^\dagger_ib_j+ U\sum_i b^\dagger_ib^\dagger_ib_ib_i$, where $b_i$ is the bosonic annihilation operator for an atom localised at site $i$, $J$ parameterises the rate of atomic hopping between neighbouring sites, and $U$ is the strength of on-site interactions between atoms.

\begin{figure}
\centering
\includegraphics[width=\linewidth]{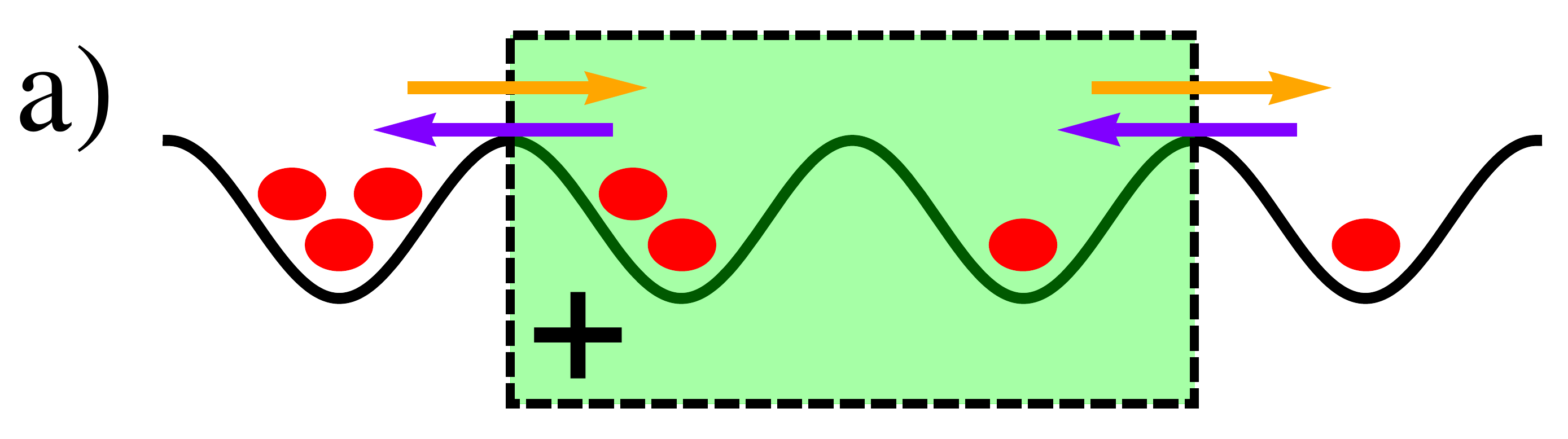}
\includegraphics[width=\linewidth]{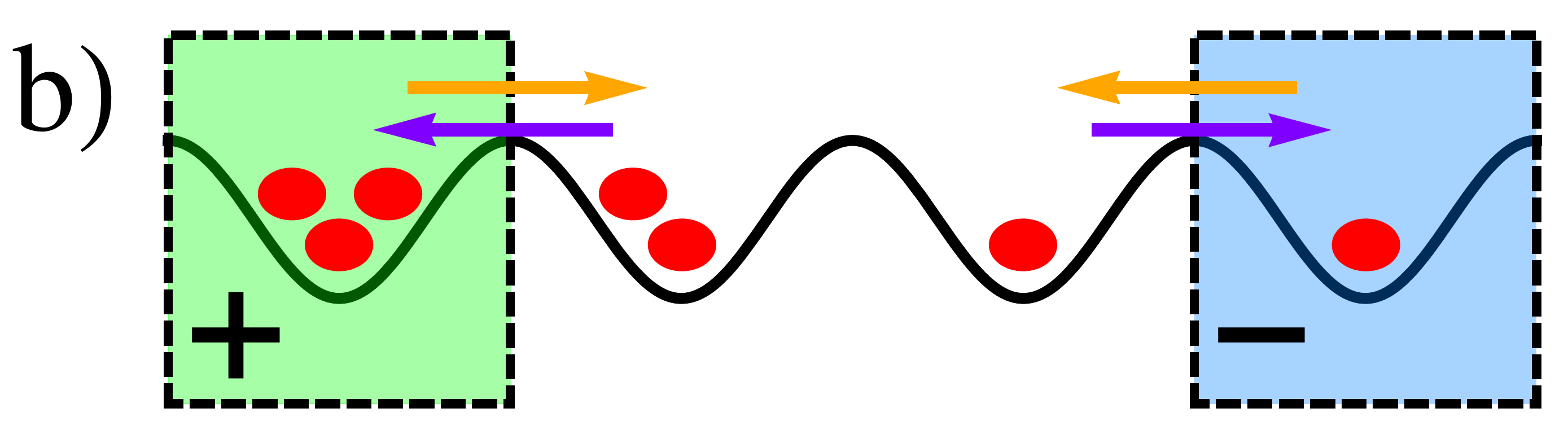}
\caption{{\bf Correlated atomic processes}: (a) Frequent, consistent measurement of the occupation of a set of sites facilitates correlated tunnelling of atoms over the boundaries of the measured region, while (b) measuring occupation number differences gives rise to pair process-like effects. Coloured boxes indicate the measured regions (green positive, blue negative), and arrows the same colour indicate correlated tunnelling events.}
\label{figatoms}
\end{figure}

Measurement of functions of atomic occupation numbers of lattice sites controls the allowed tunnelling processes $b^\dagger_i b_j$, forbidding those that change the measurement value, and correlating sets of tunnelling events that together preserve it (analogous effects occur for continuous measurement with quantum jumps \cite{mazzucchi2015}). In \figref{figatoms}(a) we illustrate how measuring the total occupation of a central region mediates long-range correlated tunnelling events across the boundaries of the region. The Zeno Hamiltonian is given by (again with $\mathcal{A}$ indicating the measured sites, and $\mathcal{B}$ the unmeasured)
\begin{equation}
\label{eqatomsaddone}
H_Z^{(1)}=-J(\sum_{\langle i\in\mathcal{A},j\in\mathcal{A}\rangle}b^\dagger_ib_j+\sum_{\langle i\in\mathcal{B},j\in\mathcal{B}\rangle}b^\dagger_ib_j)+U\sum_ib^\dagger_ib_i^\dagger b_ib_i,
\end{equation}
thus allowing tunnelling between pairs of sites where either both or neither are in the measured region, and leaving the on-site interactions unaffected. The second-order quasi-Zeno Hamiltonian is given by
\begin{equation}
\label{eqatomsaddtwo}
H_Z^{(2)}=J^2\sum_{\substack{\langle i\in\mathcal{A},j\in\mathcal{B}\rangle\\ \langle k\in\mathcal{B},l\in\mathcal{A}\rangle}}(b_i^\dagger b_jb_k^\dagger b_l+b_j^\dagger b_i b_l^\dagger b_k).
\end{equation}
This mediates correlated pairs of tunnelling events between site pairs that straddle the boundaries of the two regions $\mathcal{A}$ and $\mathcal{B}$, preserving the total occupation of sites within region $\mathcal{A}$. When the two site pairs are identical (that is, $i=l$ and $j=k$), the associated terms can be re-expressed as effective chemical potential and nearest-neighbour density-density interaction terms $J^2(2n_in_j+n_i+n_j)$. In the non-interacting limit ($U=0$), this can be mapped onto to the spin chain scenario Eqs.~\eqref{eqspinsone} and \eqref{eqspinstwo} considered above, through a Holstein-Primakoff transformation \cite{mattis2006theory}.

\begin{figure*}
\centering
\includegraphics[width=\linewidth]{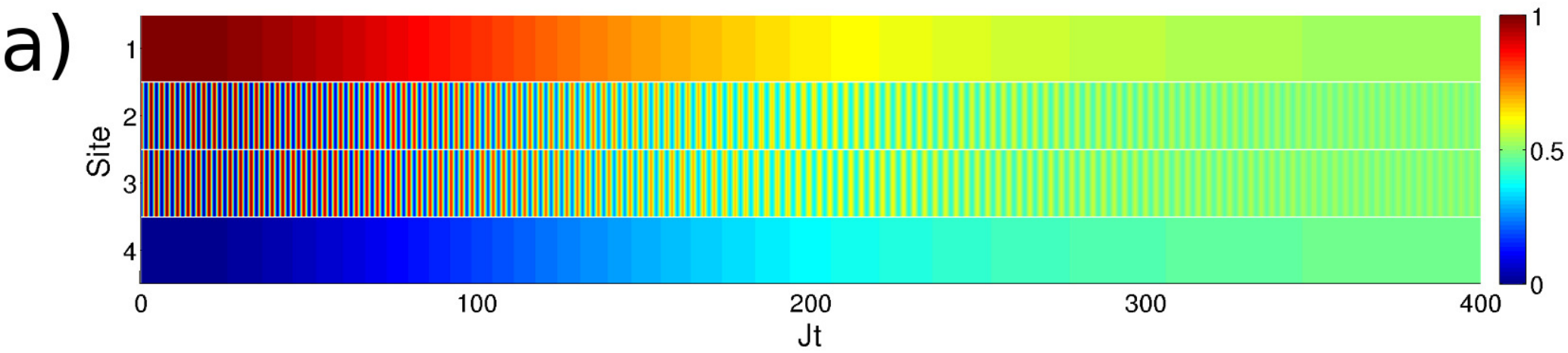}
\includegraphics[width=\linewidth]{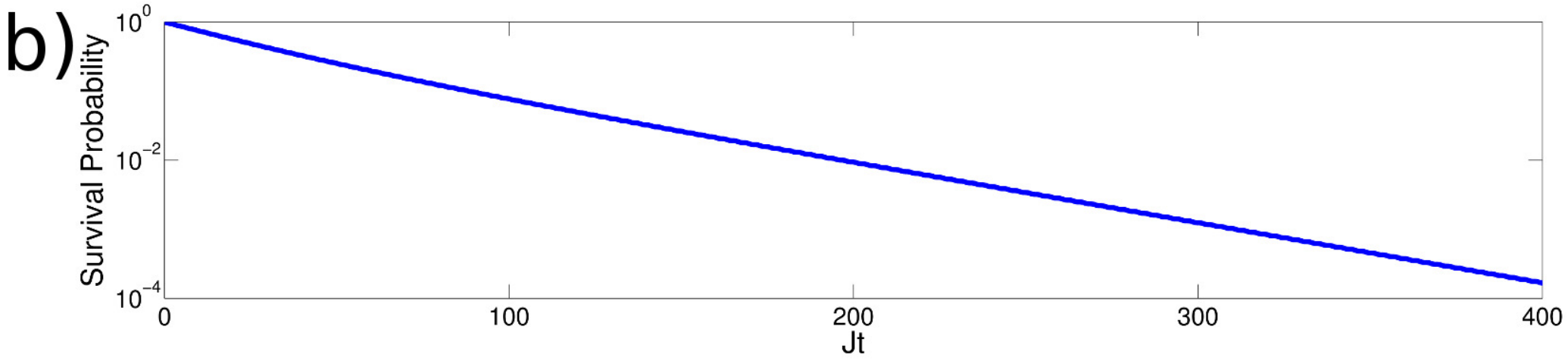}
\includegraphics[width=\linewidth]{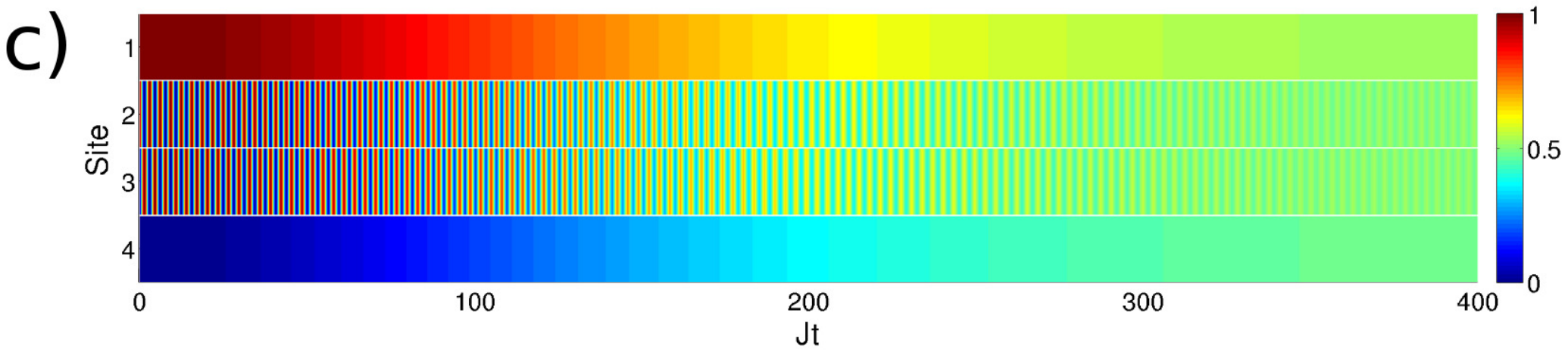}
\includegraphics[width=\linewidth]{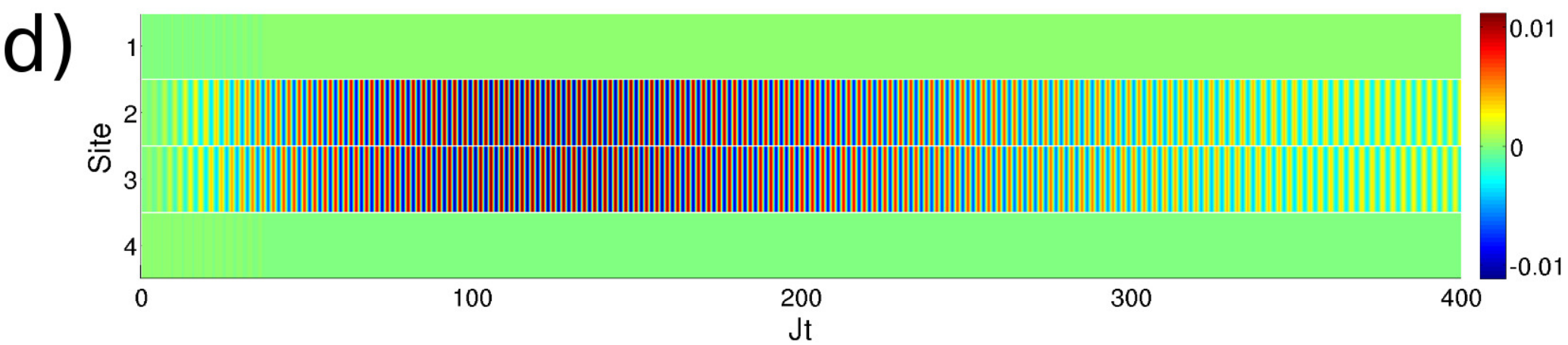}
\includegraphics[width=\linewidth]{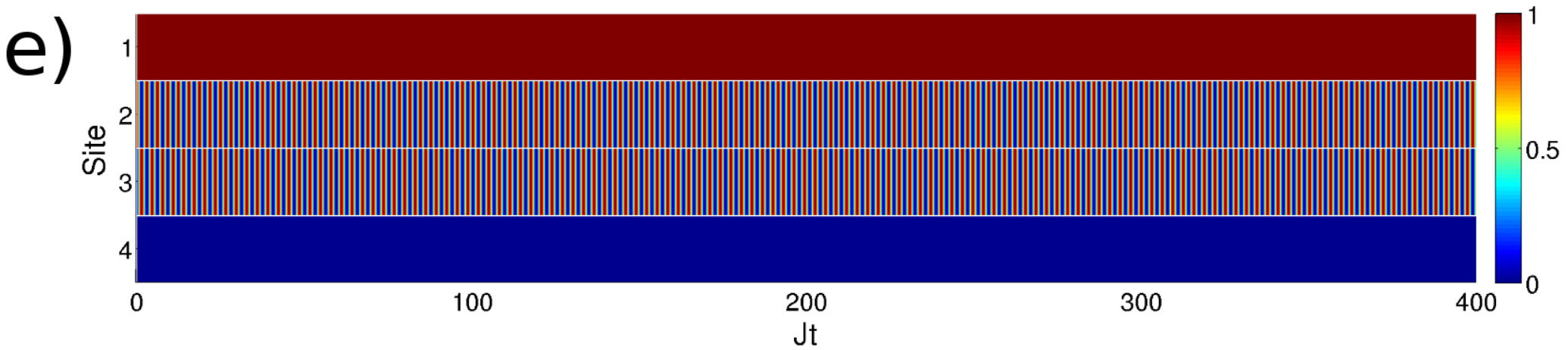}
\caption{{\bf Simulation of correlated many-body dynamics}: (a) Simulation of an effective Hamiltonian for atoms in an optical lattice, for the scheme of  Eqs.~\eqref{eqatomsaddone} and \eqref{eqatomsaddtwo} and \figref{figatoms}(a), showing first- and second-order processes. Colour map shows average site occupation. (b) Survival probability for the system to remain in the same Zeno subspace. (c) Exact evolution of the same system. (d) Difference between results for exact and effective evolution; the primary error is in capturing the first-order dynamics, and scales linearly with $\delta t$ (see main text). (e) Evolution under the Zeno Hamiltonian of QZD for the same system fails to capture the correlated processes. Simulations use 4 sites, 2 atoms, $\delta t J=10^{-2}$, $U=0$, with measurement imposing the constraint $N_2+N_3=1$. Initial state $\ket{1,1,0,0}$.}
\label{figsims}
\end{figure*}

We can also consider a scenario where the measurement is of the difference of occupation numbers at different sites. As illustrated in \figref{figatoms}(b), such a measurement scheme can give rise to correlated events resembling pair processes, where tunnelling events in to or out of a particular site/region can only occur in pairs, in order to preserve the measurement value. As before, the Zeno Hamiltonian contains the processes where tunnelling occurs between sites that are both in the same region, as well as the on-site interactions:
\begin{align}
H_Z^{(1)}=&-J(\sum_{\langle i\in\mathcal{A},j\in\mathcal{A}\rangle}b^\dagger_ib_j+\sum_{\langle i\in\mathcal{B},j\in\mathcal{B}\rangle}b^\dagger_ib_j+\sum_{\langle i\in\mathcal{C},j\in\mathcal{C}\rangle}b^\dagger_ib_j)\nonumber\\ &+U\sum_ib^\dagger_ib_i^\dagger b_ib_i,\nonumber
\end{align}
where we now label the three regions as: $\mathcal{A}$ measured (positive contribution); $\mathcal{B}$ unmeasured; and $\mathcal{C}$ measured (negative contribution). The corresponding second-order quasi-Zeno Hamiltonian is
\begin{align}
H_Z^{(2)}&=J^2(\sum_{\substack{\langle i\in\mathcal{A},j\in\mathcal{B}\rangle\\ \langle k\in\mathcal{C},l\in\mathcal{B}\rangle}}b^\dagger_ib_jb_k^\dagger b_l+\sum_{\substack{\langle i\in\mathcal{A},j\in\mathcal{B}\rangle\\ \langle k\in\mathcal{B},l\in\mathcal{A}\rangle}}b^\dagger_ib_jb_k^\dagger b_l\nonumber\\&+\sum_{\substack{\langle i\in\mathcal{B},j\in\mathcal{C}\rangle\\ \langle k\in\mathcal{C},l\in\mathcal{B}\rangle}}b^\dagger_ib_jb_k^\dagger b_l+\sum_{\substack{\langle i\in\mathcal{A},j\in\mathcal{C}\rangle\\ \langle k\in\mathcal{C},l\in\mathcal{A}\rangle}}b^\dagger_ib_jb_k^\dagger b_l+h.c.).\nonumber
\end{align}
The first term in $H_Z^{(2)}$ mediates the pair process-like effects, while the other terms correspond to the simultaneous crossing in opposite directions across the boundaries of each pairing of regions. As before, when these boundary pairs are at the same location, this is equivalent to an effective chemical potential and nearest-neighbour density-density interaction for these boundary sites.

We use a small-scale simulation to demonstrate these effects, using the scheme of  Eqs.~\eqref{eqatomsaddone} and \eqref{eqatomsaddtwo} and \figref{figatoms}(a). We simulate this setup, with 2 atoms distributed across 4 lattice sites, with no interparticle interactions ($U=0$), and the total occupation of the central two sites measured at timesteps $J \delta t=10^{-2}$ (well within the Zeno-locking regime), fixing their occupation at 1 atoms: $N_2+N_3=1$ (see Appendix \ref{secsimulation} for further details). We calculate the evolution of the system for the QqZD effective Hamiltonian, the exact evolution, and the standard QZD evolution on a trajectory where the measurement outcome is unchanging [\figref{figsims}]. We see that there is a very close agreement between the effective (a) and exact evolution (c), as shown by their difference (d). They exhibit the tunnelling between the central sites (as per standard QZD), as well as the transfer of atoms mediated by the quasi-Zeno dynamics between the two outer sites due to correlated tunnelling, and the convergence to a (set of) steady state(s). In contrast, the standard QZD evolution (e) completely fails to capture the correlated tunnelling and convergence to a steady state, showing only the tunnelling between the central sites. We also show (b) the survival probability for the system to remain in the Zeno subspace; we see that while the full convergence to the steady state takes a long time (with such trajectories occurring with low probability), the additional quasi-Zeno dynamics can still take place on timescales for which Zeno-locking is maintained with a high probability. The lower probability to reach the steady state (in comparison to the three-state example given in Section \ref{secinterp}) is in part because of the competition between the first- and second-order dynamics, as the system can be in states for which (second-order) quasi-Zeno dynamics do not take place (e.g.~$\ket{1,1,0,0}$), but leakage from the measurement subspace still can; the standard Zeno dynamics causes transitions between such states and the states for which the quasi-Zeno dynamics do take place (e.g.~$\ket{1,0,1,0}$).

The primary quantitative disagreement between the exact and approximate effective evolution is in capturing the first-order processes. This is because of the discrepancy between the exact binomial series, and the approximate effective exponential power series. Naively, one can argue this error to be $\mathcal{O}({H_Z^{(1)}}^2t\delta t)$ (up to a maximum of the largest possible occupation of the site), because for each quasi-Zeno Hamiltonian the discrepancy is in its associated second-order term in the evolution, thus making the Zeno Hamiltonian $H_Z^{(1)}$ responsible for the primary level of error. However, the convergence to a steady-state supresses the dynamics, and so curtails this error to some maximum value due to the damping of the accessible state space in time by $H_Z^{(2)}$.

\section{Discussion}
We have shown how, beyond the freezing of the observed value of a frequently repeated measurement of a system manifest by QZE/QZD, it is possible for dynamics to still take place across different states in the measurement subspace, even in the absence of direct transitions between them, via higher-order virtual processes that arise through transitions that take the system temporarily out of the measurement subspace, without altering the consistent outcome of the measurement value. We developed this QqZD formalism, and derived effective Hamiltonians to describe the system evolution. We generalised to incorporate measurements with non-equal timesteps and time-dependent Hamiltonians, and how the state and evolution of the system change when the measurement value changes. We showed that this regime generates correlated dynamics in many-body systems.

Whilst being relatively simple both mathematically and conceptually, this regime has previously been largely unexplored, despite the abundance of possibilities for which it lays the foundations. The field of dissipative dynamics, where the interactions between a system and its environment can be exploited to manipulate the dynamics of a system, and to prepare particular states of the system, has seen a lot of interest \cite{beige2000, verstraete2009, yi2012, everest2014, stannigel2014}, as has the very related field of using designed measurement as the source of dissipation for quantum system engineering \cite{mekhov2009, pedersen2014, elliott2015a, elliott2015b, mazzucchi2015b}. This work extends these ideas, as by eliminating particular processes at first-order only, whilst preserving them at second-order (or higher), can lead to the emergence of correlated dynamics, as demonstrated here. An experimental realisation of this regime would potentially be less taxing than similar experiments of standard QZE and QZD, as the requirement on the time between measurements is less stringent. The possible obstacles we foresee are the need to maintain the coherence of the system for sufficiently long times to witness the higher-order effects, and that for verification of these effects, a second observable must be measured at the start and end of the protocol that can distinguish states in the measurement subspace.

\section*{Acknowledgements}
We thank Wojciech Kozlowski and Igor Mekhov for discussions. TJE is funded by an EPSRC DTA. VV acknowledges funding from the National Research Foundation (Singapore), the Ministry of Education (Singapore), EPSRC, the Templeton Foundation, the Leverhulme Trust, the Oxford Martin School, and Wolfson College, University of Oxford. 

\appendix

\section{Derivation of Quantum quasi-Zeno Dynamics}
\label{secderivation}

Here we provide details on the derivation of QqZD, and clarify assumptions made about the system evolution. Firstly, we justify modeling the system as evolving under unitary evolution between measurements. This is true in general for an isolated, closed quantum system. Appealing to the so-called church of higher Hilbert space (CHHS) \cite{vedral2006introduction}, an ancilla can be appended to the system to account for the effect of an environment, such that the total system-ancilla evolution is unitary, even if the system dynamics alone is not. If the measurement outcome depends only on the system state, and is independent of the ancilla state, then the inclusion of the ancilla does not affect the QqZD result - one has simply to trace out the ancilla from the QqZD evolution in the same manner as usual for recovering the system dynamics from a CHHS treatment. A second simplification made in our treatment is that we treat the measurement as von Neumann projections. This is in keeping with the simple derivations of QZE and QZD \cite{misra1977, facchi2008}, which have subsequently been extended to more general settings, including coupling to external `measurement devices' \cite{ruseckas2001}. These treatments that incorporate the measurement device recover QZE when the time taken for the external device to measure the system state is much shorter than the system dynamics. It has also been shown that even when the measurements are not perfectly projective, QZE can still persist \cite{layden2015}. Thus, we expect when these realistic concerns are incorporated into our simplified picture of measurement, the results should be preserved.

In deriving the effective evolution for QqZD, we make use of some important properties of projectors; they are idempotent and mutually orthogonal ($\PP_j\PP_k=\PP_j\delta_{jk}$), and together span the entire Hilbert space ($\sum_j \PP_j=\ident$) \cite{nielsen2000quantum}. 

As noted in the main text, the effect on state $\rho$ of unitary evolution followed by a projective measurement is described by $\rho\to\PP U(\delta t)\rho U^\dagger(\delta t)\PP$. Defining $U_1(\delta t) = \PP U(\delta t)$, this can be written $\rho\to U_1(\delta t)\rho U_1^\dagger(\delta t)$. Following $N$ such sets of evolution and measurement in a total time $\tau=N\delta t$, with each measurement outcome in the same subspace $\PP$, we can describe the resulting system evolution by 
$$\rho\to U_N(\delta t)\rho U^\dagger_N(\delta t),$$
 where $U_N(\delta t)=U_1(\delta t)^N=(\PP U(\delta t))^N$. Expanding $U(\delta t)$ as a power series in terms of the Hamiltonian $H$, we hence have
\begin{equation}
\label{eqexact}
U_N(\delta t)=\left(\PP \left(1 - i H \delta t - H^2 \frac{\delta t^2}{2}+\mathcal{O}(\delta t^3)\right)\right)^N.
\end{equation}

In the full QZD limit, where $\delta t\to0$ and $N\to\infty$, this binomial expansion is exactly equal to the exponential of the Zeno Hamiltonian $U_N(\tau)\to\exp(-iH_Z^{(1)}\tau)$, giving the standard QZD result. Close to, but outside of this limit, we can still approximately describe this evolution by an exponential, but with the inclusion of the higher-order terms added perturbatively. These corrections can be found by examining the difference between the exact evolution, and the evolution under the Zeno Hamiltonian $H_Z^{(1)}$. Specifically, focussing on the $\mathcal{O}(\delta t^2)$ term in the power series expansion, we see that the exact evolution contains the term $-\PP H^2 \PP\delta t^2/2$, whereas an expansion of the exponential of the Zeno Hamiltonian yields $-{H_Z^{(1)}}^2\delta t^2/2$. We use the difference between these two to motivate our definition of the second-order quasi-Zeno Hamiltonian: $H_Z^{(2)}=\PP H (\mathbb{I}-\PP) H \PP$, and the requisite correction is given by $-H_Z^{(2)}\delta t^2/2$. Moreover, we then define the general quasi-Zeno Hamiltonian as $H_Z^{(k)}=\PP H ((\mathbb{I}-\PP)H)^{k-1}\PP$ to account for the corrections at higher orders of $\delta t$. These corrections then lead to the definition of the effective Hamiltonian, which contains the QqZD corrections to the Zeno Hamiltonian:
$$H_{\mathrm{eff}}=\sum_{k=1}^\infty\frac{(-i\delta t)^{k-1}}{k!}H_Z^{(k)},$$
as previously stated in Eq.~\eqref{eqHeff}. Taking this corrected effective Hamiltonian, we proceed as with the standard QZD case and exponentiate it to give the effective evolution operator $U_{\mathrm{eff}}(\tau)=\exp(-iH_{\mathrm{eff}}\tau)$, and the associated system evolution at time $\tau=N\delta t$ after $N$ measurements is hence
$$\rho\to U_{\mathrm{eff}}(\tau)\rho U_{\mathrm{eff}}^\dagger(\tau).$$
We note that this replacement of the binomial series by an exponential is approximate, and becomes exact only in the limit $\delta t\to0$ (for fixed $\tau$). However, as we are in the limit $|H \delta t|\ll1$, this approximation is still very faithful to the exact evolution, as can be witnessed in our simulation.

The $k$th-order quasi-Zeno Hamiltonian mediates $k$th-order transitions, where the initial and final states are in the measurement subspace $\mathcal{P}$, while all the intermediate states are not. The absence of processes in these Hamiltonians which have intermediate return to the measurement subspace is due to such terms already arising from products of lower-order quasi-Zeno Hamiltonians; as the higher-order quasi-Zeno Hamiltonians are intended as corrections to the evolution described by the lower-order quasi-Zeno Hamiltonians this is not surprising. For example, the term $\PP H \PP H \PP$ can be obtained from ${H_Z^{(1)}}^2$, while $\PP H (\mathbb{I}-\PP)H \PP H \PP=H_Z^{(2)}H_Z^{(1)}$.

\section{Simulation Details}
\label{secsimulation}

As stated in the main text, we simulate the scenario of  Eqs.~\eqref{eqatomsaddone} and \eqref{eqatomsaddtwo} and \figref{figatoms}(a) in \figref{figsims}, for 2 atoms distributed across 4 lattice sites. The chosen parameters are $J\delta t=10^{-2}$ and $U=0$, with measurement of the central two sites fixing $N_2+N_3=1$, and initial state $\ket{1,1,0,0}$. We take the outcome of each measurement to be consistent with this value; that is, we post-select the trajectory in which there are no jumps to other subspaces. 

Applying this specific case to Hamiltonians Eqs.~\eqref{eqatomsaddone} and \eqref{eqatomsaddtwo}, the Zeno Hamiltonian contains only the tunnelling terms between sites 2 and 3:
\begin{equation}
\label{eqsimone}
H_Z^{(1)}=-J(b^\dagger_2b_3+b^\dagger_3b_2).
\end{equation}
The second-order quasi-Zeno Hamiltonian contains terms where atoms tunnel $1\to2$ and $3\to4$ in a correlated manner (and the reverse process), as well as correlated tunnellings across the same barrier in opposite directions:
\begin{align}
H_Z^{(2)}&=J^2(2(b_1^\dagger b_2b_3^\dagger b_4 + b_2^\dagger b_1 b_4^\dagger b_3)\nonumber \\
&+b_1^\dagger b_2 b_2^\dagger b_1 + b_2^\dagger b_1 b_1^\dagger b_2 \nonumber \\
&+ b_3^\dagger b_4 b_4^\dagger b_3 +b_4^\dagger b_3 b_3^\dagger b_4).\nonumber 
\end{align}

This can be rearranged and rewritten in terms of the effective chemical potentials and nearest-neighbour density-density interactions:
\begin{align}
\label{eqsimtwo}
H_Z^{(2)}&=J^2(2(b_1^\dagger b_2b_3^\dagger b_4 + b_2^\dagger b_1 b_4^\dagger b_3)\nonumber \\
&+2(n_1n_2+n_3n_4)+n_1+n_2+n_3+n_4).
\end{align}

The constraint imposed by having the central two sites' occupation fixed at 1 atom reduces the size of the accessible state space. There are 2 possible states accessible to these two sites ($\ket{1,0}$ and $\ket{0,1}$). Similarly, given that the total number of atoms is also fixed at 2, the outer two sites 1 and 4 also have their occupation fixed at 1 atom, and thus may be any of the same 2 states, giving a total state space of dimension 4 for the whole system. This is then very amenable to exact calculations. We calculate the effective and exact evolution by using Eqs.~\eqref{eqsimone} and \eqref{eqsimtwo} with Eqs.~\eqref{eqHeff} and \eqref{eqexact} respectively, and the standard Zeno evolution by inputting Eq.~\eqref{eqsimone} into $U_{\mathrm{QZD}}(\tau)=\exp(-iH_Z^{(1)}\tau)$.


\begin{thebibliography}{10}
\expandafter\ifx\csname url\endcsname\relax
  \def\url#1{\texttt{#1}}\fi
\expandafter\ifx\csname urlprefix\endcsname\relax\def\urlprefix{URL }\fi
\providecommand{\bibinfo}[2]{#2}
\providecommand{\eprint}[2][]{\url{#2}}

\bibitem{misra1977}
\bibinfo{author}{Misra, B.} \& \bibinfo{author}{Sudarshan, E. C.~G.}
\newblock \bibinfo{title}{The {Zeno's} paradox in quantum theory}.
\newblock \emph{\bibinfo{journal}{J. Math. Phys.}}
  \textbf{\bibinfo{volume}{18}}, \bibinfo{pages}{756--763}
  (\bibinfo{year}{1977}).

\bibitem{neumann1955mathematical}
\bibinfo{author}{von Neumann, J.}
\newblock \emph{\bibinfo{title}{Mathematical Foundations of Quantum
  Mechanics}}.
\newblock Investigations in physics (\bibinfo{publisher}{Princeton University
  Press}, \bibinfo{year}{1955}).

\bibitem{teuscher2004}
\bibinfo{author}{Teuscher, C.}
\newblock \emph{\bibinfo{title}{{Alan Turing:} Life and Legacy of a Great
  Thinker}} (\bibinfo{publisher}{Springer}, \bibinfo{year}{2004}).

\bibitem{facchi2000}
\bibinfo{author}{Facchi, P.}, \bibinfo{author}{Gorini, V.},
  \bibinfo{author}{Marmo, G.}, \bibinfo{author}{Pascazio, S.} \&
  \bibinfo{author}{Sudarshan, E.}
\newblock \bibinfo{title}{Quantum {Zeno} dynamics}.
\newblock \emph{\bibinfo{journal}{Phys. Lett. A}}
  \textbf{\bibinfo{volume}{275}}, \bibinfo{pages}{12--19}
  (\bibinfo{year}{2000}).

\bibitem{facchi2002}
\bibinfo{author}{Facchi, P.} \& \bibinfo{author}{Pascazio, S.}
\newblock \bibinfo{title}{Quantum {Zeno} subspaces}.
\newblock \emph{\bibinfo{journal}{Phys. Rev. Lett.}}
  \textbf{\bibinfo{volume}{89}}, \bibinfo{pages}{080401}
  (\bibinfo{year}{2002}).

\bibitem{facchi2008}
\bibinfo{author}{Facchi, P.} \& \bibinfo{author}{Pascazio, S.}
\newblock \bibinfo{title}{Quantum {Zeno} dynamics: mathematical and physical
  aspects}.
\newblock \emph{\bibinfo{journal}{J. Phys. A}} \textbf{\bibinfo{volume}{41}},
  \bibinfo{pages}{493001} (\bibinfo{year}{2008}).

\bibitem{itano1990}
\bibinfo{author}{Itano, W.~M.}, \bibinfo{author}{Heinzen, D.~J.},
  \bibinfo{author}{Bollinger, J.~J.} \& \bibinfo{author}{Wineland, D.~J.}
\newblock \bibinfo{title}{Quantum {Zeno} effect}.
\newblock \emph{\bibinfo{journal}{Phys. Rev. A}} \textbf{\bibinfo{volume}{41}},
  \bibinfo{pages}{2295--2300} (\bibinfo{year}{1990}).

\bibitem{kwiat1995}
\bibinfo{author}{Kwiat, P.}, \bibinfo{author}{Weinfurter, H.},
  \bibinfo{author}{Herzog, T.}, \bibinfo{author}{Zeilinger, A.} \&
  \bibinfo{author}{Kasevich, M.~A.}
\newblock \bibinfo{title}{Interaction-free measurement}.
\newblock \emph{\bibinfo{journal}{Phys. Rev. Lett.}}
  \textbf{\bibinfo{volume}{74}}, \bibinfo{pages}{4763--4766}
  (\bibinfo{year}{1995}).

\bibitem{xiao2006}
\bibinfo{author}{Xiao, L.} \& \bibinfo{author}{Jones, J.~A.}
\newblock \bibinfo{title}{NMR analogues of the quantum {Zeno} effect}.
\newblock \emph{\bibinfo{journal}{Phys. Lett. A}}
  \textbf{\bibinfo{volume}{359}}, \bibinfo{pages}{424 -- 427}
  (\bibinfo{year}{2006}).

\bibitem{signoles2014}
\bibinfo{author}{Signoles, A.} \emph{et~al.}
\newblock \bibinfo{title}{Confined quantum {Zeno} dynamics of a watched atomic
  arrow}.
\newblock \emph{\bibinfo{journal}{Nat. Phys.}} \textbf{\bibinfo{volume}{10}},
  \bibinfo{pages}{715--719} (\bibinfo{year}{2014}).

\bibitem{streed2006}
\bibinfo{author}{Streed, E.~W.} \emph{et~al.}
\newblock \bibinfo{title}{Continuous and pulsed quantum {Zeno} effect}.
\newblock \emph{\bibinfo{journal}{Phys. Rev. Lett.}}
  \textbf{\bibinfo{volume}{97}}, \bibinfo{pages}{260402}
  (\bibinfo{year}{2006}).

\bibitem{schafer2014}
\bibinfo{author}{Sch{\"a}fer, F.} \emph{et~al.}
\newblock \bibinfo{title}{Experimental realization of quantum {Zeno} dynamics}.
\newblock \emph{\bibinfo{journal}{Nat. Comm.}} \textbf{\bibinfo{volume}{5}},
  \bibinfo{pages}{3194} (\bibinfo{year}{2014}).

\bibitem{patil2015}
\bibinfo{author}{Patil, Y.~S.}, \bibinfo{author}{Chakram, S.} \&
  \bibinfo{author}{Vengalattore, M.}
\newblock \bibinfo{title}{Measurement-induced localization of an ultracold
  lattice gas}.
\newblock \emph{\bibinfo{journal}{Phys. Rev. Lett.}}
  \textbf{\bibinfo{volume}{115}}, \bibinfo{pages}{140402}
  (\bibinfo{year}{2015}).

\bibitem{nakazato2003}
\bibinfo{author}{Nakazato, H.}, \bibinfo{author}{Takazawa, T.} \&
  \bibinfo{author}{Yuasa, K.}
\newblock \bibinfo{title}{Purification through {Zeno}-like measurements}.
\newblock \emph{\bibinfo{journal}{Phys. Rev. Lett.}}
  \textbf{\bibinfo{volume}{90}}, \bibinfo{pages}{060401}
  (\bibinfo{year}{2003}).

\bibitem{nakazato2004}
\bibinfo{author}{Nakazato, H.}, \bibinfo{author}{Unoki, M.} \&
  \bibinfo{author}{Yuasa, K.}
\newblock \bibinfo{title}{Preparation and entanglement purification of qubits
  through {Zeno-like} measurements}.
\newblock \emph{\bibinfo{journal}{Phys. Rev. A}} \textbf{\bibinfo{volume}{70}},
  \bibinfo{pages}{012303} (\bibinfo{year}{2004}).

\bibitem{erez2004}
\bibinfo{author}{Erez, N.}, \bibinfo{author}{Aharonov, Y.},
  \bibinfo{author}{Reznik, B.} \& \bibinfo{author}{Vaidman, L.}
\newblock \bibinfo{title}{Correcting quantum errors with the {Zeno} effect}.
\newblock \emph{\bibinfo{journal}{Phys. Rev. A}} \textbf{\bibinfo{volume}{69}},
  \bibinfo{pages}{062315} (\bibinfo{year}{2004}).

\bibitem{wu2004}
\bibinfo{author}{Wu, L.-A.}, \bibinfo{author}{Lidar, D.~A.} \&
  \bibinfo{author}{Schneider, S.}
\newblock \bibinfo{title}{Long-range entanglement generation via frequent
  measurements}.
\newblock \emph{\bibinfo{journal}{Phys. Rev. A}} \textbf{\bibinfo{volume}{70}},
  \bibinfo{pages}{032322} (\bibinfo{year}{2004}).

\bibitem{compagno2004}
\bibinfo{author}{Compagno, G.} \emph{et~al.}
\newblock \bibinfo{title}{Distillation of entanglement between distant systems
  by repeated measurements on an entanglement mediator}.
\newblock \emph{\bibinfo{journal}{Phys. Rev. A}} \textbf{\bibinfo{volume}{70}},
  \bibinfo{pages}{052316} (\bibinfo{year}{2004}).

\bibitem{militello2004}
\bibinfo{author}{Militello, B.} \& \bibinfo{author}{Messina, A.}
\newblock \bibinfo{title}{Distilling angular momentum nonclassical states in
  trapped ions}.
\newblock \emph{\bibinfo{journal}{Phys. Rev. A}} \textbf{\bibinfo{volume}{70}},
  \bibinfo{pages}{033408} (\bibinfo{year}{2004}).

\bibitem{wang2008}
\bibinfo{author}{Wang, X.-B.}, \bibinfo{author}{You, J.~Q.} \&
  \bibinfo{author}{Nori, F.}
\newblock \bibinfo{title}{Quantum entanglement via two-qubit quantum {Zeno}
  dynamics}.
\newblock \emph{\bibinfo{journal}{Phys. Rev. A}} \textbf{\bibinfo{volume}{77}},
  \bibinfo{pages}{062339} (\bibinfo{year}{2008}).

\bibitem{erez2008}
\bibinfo{author}{Erez, N.}, \bibinfo{author}{Gordon, G.},
  \bibinfo{author}{Nest, M.} \& \bibinfo{author}{Kurizki, G.}
\newblock \bibinfo{title}{Thermodynamic control by frequent quantum
  measurements}.
\newblock \emph{\bibinfo{journal}{Nature}} \textbf{\bibinfo{volume}{452}},
  \bibinfo{pages}{724--727} (\bibinfo{year}{2008}).

\bibitem{pazsilva2012}
\bibinfo{author}{Paz-Silva, G.~A.}, \bibinfo{author}{Rezakhani, A.~T.},
  \bibinfo{author}{Dominy, J.~M.} \& \bibinfo{author}{Lidar, D.~A.}
\newblock \bibinfo{title}{{Zeno} effect for quantum computation and control}.
\newblock \emph{\bibinfo{journal}{Phys. Rev. Lett.}}
  \textbf{\bibinfo{volume}{108}}, \bibinfo{pages}{080501}
  (\bibinfo{year}{2012}).

\bibitem{raimond2012}
\bibinfo{author}{Raimond, J.~M.} \emph{et~al.}
\newblock \bibinfo{title}{Quantum {Zeno} dynamics of a field in a cavity}.
\newblock \emph{\bibinfo{journal}{Phys. Rev. A}} \textbf{\bibinfo{volume}{86}},
  \bibinfo{pages}{032120} (\bibinfo{year}{2012}).

\bibitem{burgarth2013}
\bibinfo{author}{Burgarth, D.} \emph{et~al.}
\newblock \bibinfo{title}{{Non-Abelian} phases from quantum {Zeno} dynamics}.
\newblock \emph{\bibinfo{journal}{Phys. Rev. A}} \textbf{\bibinfo{volume}{88}},
  \bibinfo{pages}{042107} (\bibinfo{year}{2013}).

\bibitem{everest2014}
\bibinfo{author}{Everest, B.}, \bibinfo{author}{Hush, M.~R.} \&
  \bibinfo{author}{Lesanovsky, I.}
\newblock \bibinfo{title}{Many-body out-of-equilibrium dynamics of hard-core
  lattice bosons with nonlocal loss}.
\newblock \emph{\bibinfo{journal}{Phys. Rev. B}} \textbf{\bibinfo{volume}{90}},
  \bibinfo{pages}{134306} (\bibinfo{year}{2014}).

\bibitem{stannigel2014}
\bibinfo{author}{Stannigel, K.} \emph{et~al.}
\newblock \bibinfo{title}{Constrained dynamics via the {Zeno} effect in quantum
  simulation: Implementing non-abelian lattice gauge theories with cold atoms}.
\newblock \emph{\bibinfo{journal}{Phys. Rev. Lett.}}
  \textbf{\bibinfo{volume}{112}}, \bibinfo{pages}{120406}
  (\bibinfo{year}{2014}).

\bibitem{elliott2015a}
\bibinfo{author}{Elliott, T.~J.}, \bibinfo{author}{Kozlowski, W.},
  \bibinfo{author}{Caballero-Benitez, S.~F.} \& \bibinfo{author}{Mekhov, I.~B.}
\newblock \bibinfo{title}{Multipartite entangled spatial modes of ultracold
  atoms generated and controlled by quantum measurement}.
\newblock \emph{\bibinfo{journal}{Phys. Rev. Lett.}}
  \textbf{\bibinfo{volume}{114}}, \bibinfo{pages}{113604}
  (\bibinfo{year}{2015}).

\bibitem{mazzucchi2015}
\bibinfo{author}{Mazzucchi, G.}, \bibinfo{author}{Kozlowski, W.},
  \bibinfo{author}{Caballero-Benitez, S.~F.}, \bibinfo{author}{Elliott, T.~J.}
  \& \bibinfo{author}{Mekhov, I.~B.}
\newblock \bibinfo{title}{Quantum measurement-induced dynamics of many-body
  ultracold bosonic and fermionic systems in optical lattices}.
\newblock \emph{\bibinfo{journal}{arXiv:1503.08710}}  (\bibinfo{year}{2015}).

\bibitem{elliott2015b}
\bibinfo{author}{Elliott, T.~J.} \& \bibinfo{author}{Mekhov, I.~B.}
\newblock \bibinfo{title}{Engineering many-body dynamics with quantum light
  potentials and measurements}.
\newblock \emph{\bibinfo{journal}{Phys. Rev. A}} \textbf{\bibinfo{volume}{94}},
  \bibinfo{pages}{013614} (\bibinfo{year}{2016}).

\bibitem{layden2015}
\bibinfo{author}{Layden, D.}, \bibinfo{author}{Mart\'{\i}n-Mart\'{\i}nez, E.}
  \& \bibinfo{author}{Kempf, A.}
\newblock \bibinfo{title}{Perfect {Zeno-like} effect through imperfect
  measurements at a finite frequency}.
\newblock \emph{\bibinfo{journal}{Phys. Rev. A}} \textbf{\bibinfo{volume}{91}},
  \bibinfo{pages}{022106} (\bibinfo{year}{2015}).

\bibitem{dhar2015}
\bibinfo{author}{Dhar, S.}, \bibinfo{author}{Dasgupta, S.},
  \bibinfo{author}{Dhar, A.} \& \bibinfo{author}{Sen, D.}
\newblock \bibinfo{title}{Detection of a quantum particle on a lattice under
  repeated projective measurements}.
\newblock \emph{\bibinfo{journal}{Phys. Rev. A}} \textbf{\bibinfo{volume}{91}},
  \bibinfo{pages}{062115} (\bibinfo{year}{2015}).

\bibitem{kozlowski2015}
\bibinfo{author}{Kozlowski, W.}, \bibinfo{author}{Caballero-Benitez, S.~F.} \&
  \bibinfo{author}{Mekhov, I.~B.}
\newblock \bibinfo{title}{{Non-Hermitian} dynamics in the quantum {Zeno}
  limit}.
\newblock \emph{\bibinfo{journal}{arXiv:1510.04857}}  (\bibinfo{year}{2015}).

\bibitem{dirac1967principles}
\bibinfo{author}{Dirac, P. A.~M.}
\newblock \emph{\bibinfo{title}{The Principles of Quantum Mechanics}}
  (\bibinfo{publisher}{Clarendon Press, Oxford}, \bibinfo{year}{1967}).

\bibitem{zee2010}
\bibinfo{author}{Zee, A.}
\newblock \emph{\bibinfo{title}{Quantum Field Theory in a Nutshell}}
  (\bibinfo{publisher}{Princeton University Press}, \bibinfo{year}{2010}).

\bibitem{mattis2006theory}
\bibinfo{author}{Mattis, D.}
\newblock \emph{\bibinfo{title}{The Theory of Magnetism Made Simple: An
  Introduction to Physical Concepts and to Some Useful Mathematical Methods}}
  (\bibinfo{publisher}{World Scientific}, \bibinfo{year}{2006}).

\bibitem{kramers1934}
\bibinfo{author}{Kramers, H.~A.}
\newblock \bibinfo{title}{{L'interaction Entre les Atomes Magnétogènes dans
  un Cristal Paramagnétique}}.
\newblock \emph{\bibinfo{journal}{Physica}} \textbf{\bibinfo{volume}{1}},
  \bibinfo{pages}{182 -- 192} (\bibinfo{year}{1934}).

\bibitem{anderson1950}
\bibinfo{author}{Anderson, P.~W.}
\newblock \bibinfo{title}{Antiferromagnetism. theory of superexchange
  interaction}.
\newblock \emph{\bibinfo{journal}{Phys. Rev.}} \textbf{\bibinfo{volume}{79}},
  \bibinfo{pages}{350--356} (\bibinfo{year}{1950}).

\bibitem{landig2015}
\bibinfo{author}{Landig, R.} \emph{et~al.}
\newblock \bibinfo{title}{Quantum phases emerging from competing short- and
  long-range interactions in an optical lattice}.
\newblock \emph{\bibinfo{journal}{arXiv:1511.00007}}  (\bibinfo{year}{2015}).

\bibitem{klinder2015}
\bibinfo{author}{Klinder, J.}, \bibinfo{author}{Ke{\ss}ler, H.},
  \bibinfo{author}{Bakhtiari, M.~R.}, \bibinfo{author}{Thorwart, M.} \&
  \bibinfo{author}{Hemmerich, A.}
\newblock \bibinfo{title}{Observation of a superradiant {Mott} insulator in the
  {Dicke-Hubbard} model}.
\newblock \emph{\bibinfo{journal}{arXiv:1511.00850}}  (\bibinfo{year}{2015}).

\bibitem{mekhov2007}
\bibinfo{author}{Mekhov, I.~B.}, \bibinfo{author}{Maschler, C.} \&
  \bibinfo{author}{Ritsch, H.}
\newblock \bibinfo{title}{Probing quantum phases of ultracold atoms in optical
  lattices by transmission spectra in cavity quantum electrodynamics}.
\newblock \emph{\bibinfo{journal}{Nat. Phys.}} \textbf{\bibinfo{volume}{3}},
  \bibinfo{pages}{319--323} (\bibinfo{year}{2007}).

\bibitem{jaksch1998}
\bibinfo{author}{Jaksch, D.}, \bibinfo{author}{Bruder, C.},
  \bibinfo{author}{Cirac, J.~I.}, \bibinfo{author}{Gardiner, C.~W.} \&
  \bibinfo{author}{Zoller, P.}
\newblock \bibinfo{title}{Cold bosonic atoms in optical lattices}.
\newblock \emph{\bibinfo{journal}{Phys. Rev. Lett.}}
  \textbf{\bibinfo{volume}{81}}, \bibinfo{pages}{3108--3111}
  (\bibinfo{year}{1998}).

\bibitem{beige2000}
\bibinfo{author}{Beige, A.}, \bibinfo{author}{Braun, D.},
  \bibinfo{author}{Tregenna, B.} \& \bibinfo{author}{Knight, P.~L.}
\newblock \bibinfo{title}{Quantum computing using dissipation to remain in a
  decoherence-free subspace}.
\newblock \emph{\bibinfo{journal}{Phys. Rev. Lett.}}
  \textbf{\bibinfo{volume}{85}}, \bibinfo{pages}{1762} (\bibinfo{year}{2000}).

\bibitem{verstraete2009}
\bibinfo{author}{Verstraete, F.}, \bibinfo{author}{Wolf, M.~M.} \&
  \bibinfo{author}{Cirac, J.~I.}
\newblock \bibinfo{title}{Quantum computation and quantum-state engineering
  driven by dissipation}.
\newblock \emph{\bibinfo{journal}{Nat. Phys.}} \textbf{\bibinfo{volume}{5}},
  \bibinfo{pages}{633--636} (\bibinfo{year}{2009}).

\bibitem{yi2012}
\bibinfo{author}{Yi, W.}, \bibinfo{author}{Diehl, S.}, \bibinfo{author}{Daley,
  A.~J.} \& \bibinfo{author}{Zoller, P.}
\newblock \bibinfo{title}{Driven-dissipative many-body pairing states for cold
  fermionic atoms in an optical lattice}.
\newblock \emph{\bibinfo{journal}{New Journal of Physics}}
  \textbf{\bibinfo{volume}{14}}, \bibinfo{pages}{055002}
  (\bibinfo{year}{2012}).

\bibitem{mekhov2009}
\bibinfo{author}{Mekhov, I.~B.} \& \bibinfo{author}{Ritsch, H.}
\newblock \bibinfo{title}{Quantum optics with quantum gases: Controlled state
  reduction by designed light scattering}.
\newblock \emph{\bibinfo{journal}{Phys. Rev. A}} \textbf{\bibinfo{volume}{80}},
  \bibinfo{pages}{013604} (\bibinfo{year}{2009}).

\bibitem{pedersen2014}
\bibinfo{author}{Pedersen, M.~K.}, \bibinfo{author}{S{\o}rensen, J. J.~W.},
  \bibinfo{author}{Tichy, M.~C.} \& \bibinfo{author}{Sherson, J.~F.}
\newblock \bibinfo{title}{Many-body state engineering using measurements and
  fixed unitary dynamics}.
\newblock \emph{\bibinfo{journal}{New J. Phys.}} \textbf{\bibinfo{volume}{16}},
  \bibinfo{pages}{113038} (\bibinfo{year}{2014}).

\bibitem{mazzucchi2015b}
\bibinfo{author}{Mazzucchi, G.}, \bibinfo{author}{Caballero-Benitez, S.~F.} \&
  \bibinfo{author}{Mekhov, I.~B.}
\newblock \bibinfo{title}{Quantum measurement-induced antiferromagnetic order
  and density modulations in ultracold fermi gases in optical lattices}.
\newblock \emph{\bibinfo{journal}{arXiv:1510.04883}}  (\bibinfo{year}{2015}).

\bibitem{vedral2006introduction}
\bibinfo{author}{Vedral, V.}
\newblock \emph{\bibinfo{title}{Introduction to Quantum Information Science}}
  (\bibinfo{publisher}{OUP Oxford}, \bibinfo{year}{2006}).

\bibitem{ruseckas2001}
\bibinfo{author}{Ruseckas, J.} \& \bibinfo{author}{Kaulakys, B.}
\newblock \bibinfo{title}{Real measurements and the quantum {Zeno} effect}.
\newblock \emph{\bibinfo{journal}{Phys. Rev. A}} \textbf{\bibinfo{volume}{63}},
  \bibinfo{pages}{062103} (\bibinfo{year}{2001}).

\bibitem{nielsen2000quantum}
\bibinfo{author}{Nielsen, M.} \& \bibinfo{author}{Chuang, I.}
\newblock \emph{\bibinfo{title}{Quantum Computation and Quantum Information}}
  (\bibinfo{publisher}{Cambridge University Press}, \bibinfo{year}{2000}).

\end{thebibliography}
\end{document}